\documentclass[12pt,a4paper]{article}
\usepackage{jheppub}
\usepackage{algorithm2e}
\usepackage{amsmath}
\usepackage{amssymb}
\usepackage{dsfont}
\usepackage{graphicx}
\usepackage{mathtools}
\usepackage{physics}
\usepackage{slashed}
\usepackage[usenames,dvipsnames]{xcolor}
\usepackage{color}
\definecolor{darkblue}{rgb}{0.1,0.1,.7}
\newcommand{\cG}{\hat{\mathcal G}}
\newcommand{\calO}{\mathcal{O}}
\newcommand{\calL}{\mathcal{L}}

\newcommand{\GNY}{{\rm GNY}}
\newcommand{\e}{\epsilon}
\newcommand{\eps}{\epsilon}
\newcommand{\s}{\sigma}
\newcommand{\Dsp}{\Delta_{\sigma'}}
\newcommand{\Dep}{\Delta_{\epsilon'}}
\newcommand{\Dtx}[1]{\Delta_{\text{#1}}}

\makeatletter
\def\@fpheader{\ }
\makeatother

\title{Bounding irrelevant operators in the 3d Gross--Neveu--Yukawa CFTs}
\author{Matthew S. Mitchell and David Poland}
\affiliation{Department of Physics, Yale University, 217 Prospect St, New Haven, CT 06520, USA}

\emailAdd{matthew.mitchell@yale.edu}
\emailAdd{david.poland@yale.edu}

\date{}
\abstract{
  We perform a numerical bootstrap study of scalar operators in the critical 3d Gross--Neveu--Yukawa
  models, a family of conformal field theories containing $N$ Majorana fermions in the fundamental
  representation of an $O(N)$ global symmetry. We compute rigorous bounds on the scaling dimensions
  of the next-to-lowest parity-even and parity-odd singlet scalars at $N = 2, 4,$ and $8$. All of
  these dimensions have lower bounds greater than 3, implying that there are only two relevant
  singlet scalars and placing constraints on the RG flow structure of these theories.}

\begin{document}

\maketitle

\section{Introduction}

The numerical conformal bootstrap~\cite{Rattazzi:2008pe, Poland:2018epd, Rychkov:2023wsd} is a method to rigorously bound the parameters of a conformal field theory (CFT) based on symmetry, unitarity, and assumptions about the spectrum of operators. This can be accomplish by reformulating the CFT's self-consistency constraints as a semidefinite programming problem \cite{Poland:2011ey, Simmons-Duffin:2015qma, Landry:2019qug}, which can be solved with well-understood numerical techniques. This approach has been highly successful, and has been used to obtain small ``islands'' of allowed
parameter space for a wide range of theories, most notably the Ising~\cite{ElShowk:2012ht,El-Showk:2014dwa,Kos:2014bka,Simmons-Duffin:2015qma,Kos:2016ysd,Simmons-Duffin:2016wlq} and scalar $O(N)$ CFTs~\cite{Kos:2013tga,Kos:2015mba,Kos:2016ysd,Chester:2019ifh,Chester:2020iyt}. Because CFTs correspond to critical points of condensed matter systems, the bootstrap has led to several insights about phase transitions, including the instability of Heisenberg magnets to anisotropic perturbations~\cite{Chester:2020iyt}. More recently, the bootstrap has been extended to 3d CFTs containing strongly-interacting fermions~\cite{Iliesiu:2015qra,Iliesiu:2017nrv, Erramilli:2022kgp}. 

Gross--Neveu--Yukawa (GNY) models, a family of QFTs which contain an $N$-component fermion field coupled to a scalar, are among the simplest fermionic quantum field theories exhibiting critical behavior. While there are a number of variants of these models, the simplest version is described perturbatively by the Lagrangian
\begin{equation} \label{eq:lagrangian} \calL_\GNY = -\frac12 (\partial \phi)^2 - i \frac{1}{2}
  \psi_i \slashed{\partial} \psi_i -\frac{1}{2}m^2\phi^2 -\frac{\lambda}{4}\phi^4 - i \frac{g}{2}
  \phi \psi_i\psi_i\,,
\end{equation}
which is believed to flow to an interacting CFT at a critical value of $m^2$ at all values of $N$. Here, $\psi_i$ denotes Majorana spinors with a fundamental $O(N)$ index, and the product of two $\psi$'s denotes the spinor contraction $\Omega_{\alpha\beta}\psi_i^\alpha \psi_j^\beta$.

GNY models and their variations have been proposed to describe a variety of quantum phase transitions in strongly-correlated materials with emergent Lorentz symmetry, including transitions in graphene~\cite{herbut2006interactions,Herbut:2009qb, herbut2009relativistic, Mihaila:2017ble}, d-wave superconductors~\cite{vojta2000quantum, vojta2003quantum}, and edge-modes of topological superconductors~\cite{Grover:2013rc,Ziegler:2021yua}. Further details about the applications of specific GNY models are described in section 2.5 of~\cite{Erramilli:2022kgp}.

Before proceeding further, we should clarify our spinor conventions: In $(4-\epsilon)$d, GNY models are generally taken to have $N_D$ 4-component \emph{Dirac} fermions, which transform in the fundamental representation of $U(N_D)$. In 3d, however, the 4-component Dirac representation factorizes into 4 2-component \emph{Majorana} representations, corresponding to a symmetry enhancement from $U(N_D)$ to $O(4N_D)$. In this paper, $N$ will always denote the number of Majorana fermions (that is, $N=4N_D$).

One particularly important property of these critical points is the number of relevant scalar operators, which corresponds to the codimension of the critical point in phase space---i.e.~the number of independent parameters which must be tuned to reach it. The number of parameters may be lower in a given microscopic realization, depending on which symmetries it preserves, but the number of relevant scalars provides an upper bound. In this work, we will compute two-sided bounds on the subleading operators $\sigma' \sim \phi^3$ and $\epsilon' \sim \phi \partial^2 \phi$ using the numerical bootstrap, showing that the GNY models at $N=2$, $4$, and $8$ have exactly two relevant $O(N)$-invariant scalars, one parity even ($\epsilon \sim \phi^2$) and one parity odd ($\sigma \sim \phi$). This will be similar in spirit to the study~\cite{Reehorst:2021hmp} which bounded irrelevant operators in the 3d Ising CFT.

The computation of these bounds is in part motivated by the study~\cite{Erramilli:2022kgp}, which required inputting assumptions about the spectral gaps until $\sigma'$ and $\epsilon'$. In particular, in the $N=2$ case, a significantly smaller island was obtained after assuming $\sigma'$ was irrelevant ($\Delta_{\sigma'} > 3$), in comparison with the milder assumption $\Delta_{\sigma'} > 2.5$. On the other hand, it is known that at $N=1$, the $\sigma'$ operator is relevant, while at large $N$ it is known to be irrelevant. The study in this paper was motivated by the desire to place this assumption of the irrelevance of $\sigma'$ at $N \geq 2$ on a firmer footing. 

This paper is organized as follows: in section \ref{sec:theory}, we review theoretical work on GNY CFT spectra, including large-$N$ and Padé approximations. In section \ref{sec:numerics}, we summarize our numerical setup and the algorithms that we use. In section \ref{sec:results} we describe the results of our calculations, concluding with a brief discussion of future research directions.

\section{Large-$N$ and Padé approximations} \label{sec:theory}

Essentially all conformal bootstrap studies require some physical input about the spectrum of scaling dimensions. However, one hopes to make precise predictions for low-lying data with only mild, well-motivated assumptions about gaps in the spectrum. In the 3d $O(N)$ GNY models, such gaps are easiest to justify from the perspective of either the large-$N$ or $(4-\epsilon)$ expansions. 

We summarize of our knowledge of the leading scalar operators in this theory in table~\ref{tab:large-N-and-e-exp-GNY}. In particular, we highlight that the operator $\sigma'$ is slightly irrelevant at large $N$, with dimension $\Delta_{\sigma'} = 3 + \frac{64}{\pi^2 N} - \ldots$, but with a negative $1/N^2$ correction~\cite{Manashov:2017rrx}. In contrast, the operator $\epsilon'$ is the lowest eigenvalue of the mixture between $\phi \partial^2 \phi$ and $\phi^4$, which has dimension $\Delta_{\eps'} = 4 - \frac{64}{3\pi^2 N} + \ldots$, and a positive $1/N^2$ correction~\cite{Manashov:2017rrx}. 

\begin{table}[t!]
\begin{center}
\renewcommand{\arraystretch}{1.1}
\resizebox{\columnwidth}{!}{
\begin{tabular}{l|c c|l|l }
\hline\hline Operator~& Parity & $O(N)$ & $\Delta$ at large $N$&$\Delta$ in $\e$-exp.\\
\hline $\psi_i$ & $+$ &V & $1+ \frac{4}{3\pi^2 N} + \frac{896}{27\pi^4 N^2}+ \frac{\#}{N^3} +\dots $  & $\frac{3}2 - \frac{N+5}{2(N+6)} \epsilon + \dots$ \\
 $\psi_i' \sim \phi^2 \psi_i$ & $+$ &V & $3+\frac{100}{3\pi^2 N}+ \dots$  & - \\
  $\chi_i \sim \phi^3 \psi_i$ & $-$ &V & $4+ \frac{292}{3\pi^2 N}+\dots$  & - \\
 $\sigma \sim \phi$ & $-$  &S& $1- \frac{32}{3\pi^2 N} +\frac{32 \left(304-27 \pi ^2\right)}{27 \pi ^4 N^2}+ \dots$ & $1-\frac{3}{N+6} \epsilon+ \dots$ \\
  $\sigma' \sim \phi^3 $ & $-$ &S& $3+\frac{64}{\pi^2 N} -\frac{128 \left(770-9 \pi ^2\right)}{9 \pi ^4 N^2}  + \dots$ & $3+\frac{\sqrt{N^2 +132N +36} - N - 30}{6(N+6)}\epsilon+\dots$  \\
$\sigma'' \sim \phi^5 $ & $-$ &S& $5 +\frac{800}{3 \pi ^2 N} - \frac{160(12512 - 351\pi^2)}{27 \pi^4 N^2}+ \dots$ & -\\
 $\epsilon \sim \phi^2$ & $+$ &S& $2+\frac{32}{3\pi^2 N} -\frac{64 \left(632+27 \pi ^2\right)}{27 \pi ^4 N^2}  + \dots$& $2+ \frac{\sqrt{N^2 +132 N + 36} - N-30}{6(N+6)}\epsilon + \dots$\\ 
  $\epsilon' \sim \phi \partial^2 \phi $ & $+$ &S& $4-\frac{64}{3 \pi ^2 N} + \frac{64 \left(400-27
                                                   \pi ^2\right)}{27 \pi ^4 N^2} + \dots$ & -  \\
  $\epsilon'' \sim \phi^4 $ & $+$ &S& $4+\frac{448}{3\pi^2 N} -\frac{256 \left(3520-81 \pi ^2\right)}{27 \pi ^4 N^2} + \dots$ & -  \\
  $\phi^k $ & $(-1)^k$ &S& $k +\frac{16 (3 k-5) k}{3 \pi ^2 N} - \frac{\#}{N^2}+ \dots$
& -  \\
  $\sigma_T \sim \psi_{(i}\psi_{j)} $ & $-$ &T& $2+\frac{32}{3\pi^2 N} +\frac{4096}{27 \pi ^4 N^2} +\dots$ & -\\
\hline\hline 
\end{tabular}}
\end{center}
\caption{Large-$N$ \cite{Gracey:1992cp,Derkachov:1993uw,Gracey:1993kc, Petkou:1996np, Moshe:2003xn, Iliesiu:2015qra, fei2016yukawa, Manashov:2017rrx, Gracey:2018fwq,Erramilli:2022kgp} and $\epsilon$-expansion
  estimates \cite{gracey1990three, rosenstein1993critical, zerf2016superconducting, Gracey:2016mio, fei2016yukawa, Mihaila:2017ble, Zerf:2017zqi, Ihrig:2018hho} for the leading scalar operator dimensions in the $O(N)$ GNY models. The numerators written as \(\#\) correspond to known expressions that are suppressed. 
  This table is similar to table 2 of~\cite{Erramilli:2022kgp}, but we note that our interpretation of $\epsilon'$ differs from what was presented there. In the large-$N$ expansion, the operators $\phi \partial^2 \phi$ and $\phi^4$ mix with each other, but the lower eigenvalue is predominantly $\epsilon' \sim \phi \partial^2 \phi$ and the larger eigenvalue is predominantly $\epsilon'' \sim \phi^4$~\cite{Manashov:2017rrx}.}
\label{tab:large-N-and-e-exp-GNY}
\end{table}

On the other hand, the $N=1$ GNY theory can be identified with the $\mathcal N = 1$ super-Ising model, which was bootstrapped in~\cite{Rong:2018okz, Atanasov:2018kqw, Atanasov:2022bpi}. In particular, it was determined in~\cite{Atanasov:2022bpi} that in this theory $\Delta_{\sigma} = \Delta_{\epsilon} - 1 = 0.5844435(83)$ and $\Delta_{\sigma'} = \Delta_{\epsilon'} - 1 = 2.8869(25)$. Thus, at $N=1$ the operator $\sigma'$ is known to be relevant. From the extremal spectra, this work could also (non-rigorously) extract the location of the next scalar multiplet containing $\Delta_{\epsilon''} = 4.38(1)$ and $\Delta_{\sigma''} = 5.38(1)$. 

Because the super-Ising data is so well understood, it's desirable to incorporate it into our approximations for the scaling dimensions of the GNY CFTs at larger values of $N$. We can do this by constructing Padé approximations, rational functions which match the large-$N$ expansion at $N=\infty$ and the super-Ising data at $N=1$. Similar Padé approximations were constructed previously in~\cite{Rong:2018okz, Erramilli:2022kgp}. These approximations will be plotted against our results in figures~\ref{fig:boundssig} and~\ref{fig:boundseps} and yield the estimates in table~\ref{tab:pade}.

\begin{table}[t!]
\begin{center}
\renewcommand{\arraystretch}{1.1}
\begin{tabular}{l|l l l|l l l}
  \hline\hline $N$~& $\Delta_{\sigma}$ & $\Delta_{\sigma'}$ & $\Delta_{\sigma''}$ & $\Delta_{\epsilon}$ & $\Delta_{\epsilon'}$ & $\Delta_{\epsilon''}$\\
  \hline
  1 & 0.5844435\textbf{(83)} & 2.8869\textbf{(25)} & 5.38\textbf{(1)}& 1.5844435\textbf{(83)} & 3.8869\textbf{(25)} & 4.38\textbf{(1)}\\
  2 & 0.6500\textbf{(12)} & 3.1487(11)  & 5.805(5)   & 1.725\textbf{(7)}  & 3.5474(5)  & 4.579(4)   \\
  4 & 0.7578\textbf{(15)} & 3.2455(4)   & 5.9393(19) & 1.899\textbf{(10)} & 3.63674(7) & 4.6230(18) \\
  8 & 0.8665\textbf{(13)} & 3.24739(15) & 5.8901(7)  & 2.002\textbf{(12)} & 3.77711(1) & 4.5648(6)  \\
  \hline\hline 
\end{tabular}
\end{center}
\caption{Scaling dimensions of $\{\sigma, \sigma', \sigma''\}$ and
  $\{\epsilon, \epsilon', \epsilon''\}$ in the $O(N)$ GNY models at $N=1, 2, 4, 8$. Bounds with bolded error bars are rigorous, and were obtained from the conformal bootstrap in~\cite{Erramilli:2022kgp} and~\cite{Atanasov:2022bpi}. Other estimates come from a $(2,1)$ Padé approximation, which incorporates both the large-$N$ expansion to order $1/N^2$ and the $\mathcal{N}=1$ super-Ising numerical bootstrap data~\cite{Atanasov:2022bpi} as a boundary condition at $N=1$.}
\label{tab:pade}
\end{table}

In order to compute two-sided bounds on $\sigma'$ and $\epsilon'$ we will need to impose gaps until $\sigma''$ and $\epsilon''$. The Padé estimates shown in table~\ref{tab:pade} and figures~\ref{fig:boundssig} and~\ref{fig:boundseps} motivate the conservative assumptions $\Delta_{\sigma''} > 4.5$ and $\Delta_{\epsilon''} > 4$, which will be imposed when computing bounds on $\Delta_{\sigma'}$ and $\Delta_{\epsilon'}$, respectively. 

\section{Numerical setup} \label{sec:numerics}

\subsection{Crossing equations} \label{sec:crossing}

We use an identical numerical setup to \cite{Erramilli:2022kgp}, which we shall summarize here for the reader's convenience.

We impose crossing on all non-vanishing four-point correlators of $\psi$, $\sigma$, and $\eps$. Up to permutations, these are:
\begin{equation}
  \ev{\psi \psi \psi \psi}, \ev{\eps \eps \eps \eps}, \ev{\s\s\s\s},
  \ev{\psi \psi \eps \eps}, \ev{\psi \psi \s\s }, \ev{\s \eps \psi \psi },\ev{\s\s\eps \eps }.
\end{equation}
This notation is, of course, abbreviated. In addition to the positions of the operator insertions, correlation functions depend on the operators' quantum numbers, which carry information about the global symmetry group. We can express them as:
\begin{equation}
  \ev{\calO_i(\vb p_1) \calO_j(\vb p_2) \calO_k(\vb p_3) \dots}\,,
\end{equation}
where $ijk$ label our choice of external operator ($\psi$, $\sigma$, or $\epsilon$), and $\vb p$ is a vector that encodes the position, spin, and flavor information (the latter two are generally expressed in index free notation; see e.g.~\cite{Costa:2011mg}). We can now decompose these correlators into a basis of tensor structures, which are invariant under all symmetries of the theory, and use these structures to convert the crossing equations into a semidefinite program.

Specifically, our goal is to relate the four-point structures $T_{I,ijkl}$, which appear in the expansion
\begin{equation}
  \ev{\calO_i(\vb p_1) \calO_j(\vb p_2) \calO_k(\vb p_3) \calO_l(\vb p_4)} =
  \sum_I g^I_{ijkl}(z, \bar z) T_{I,ijkl}(\vb p_1, \vb p_2, \vb p_3, \vb p_4)\,,
\end{equation}
to the three-point structures  $Q_{a,ijk}$, which appear in the expansions
\begin{align}
  \ev{\calO_i(\vb p_1) \calO_j(\vb p_2) \calO_k(\vb p_3)}
  &= \sum_a \lambda^a_{ijk} Q_{a,ijk}(\vb p_1, \vb p_2, \vb p_3 )\,,\\
  \ev{\calO_i(\vb p_1) \calO_j(\vb p_2) \calO_{\Delta,\rho}(\vb p_3)}
  &= \sum_a \lambda^a_{ij;\Delta,\rho} Q_{a,ij;\Delta,\rho}(\vb p_1, \vb p_2, \vb p_3 )\,.
\end{align}
In keeping with the conventions of \cite{Erramilli:2022kgp}, $\calO_{\Delta,\rho}$ refers to an \emph{internal} operator---one that does not appear in any of the bootstrapped four-point functions---with scaling dimension $\Delta$ and group representation $\rho \subset \rho_i \otimes \rho_j$. It is generally helpful to separate the internal and external contributions in the crossing equations. The invariant parts of the four-point function can be expanded in terms of OPE coefficients as
\begin{equation}
  g^I_{ijkl}(z,\bar z) =
  \sum_{\calO} \sum_{a,b}
  \lambda^a_{ij\calO} \lambda^b_{kl\calO}
  G_{ab,ijkl,\calO}^I(z,\bar z)\,,
\end{equation}
where $\calO$ runs over both internal and external operators (including the identity), and $G$ can be computed with the algorithm implemented in \texttt{blocks\_3d} \cite{Erramilli:2019njx, Erramilli:2020rlr}.

The crossing equations, which take the form
\begin{equation}
  \ev{\calO_i(\vb p_1) \calO_j(\vb p_2) \calO_k(\vb p_3) \calO_l(\vb p_4)} \pm 
  \ev{\calO_k(\vb p_3) \calO_j(\vb p_2) \calO_i(\vb p_1) \calO_l(\vb p_4)} = 0\,,
\end{equation}
can now be decomposed into four-point structures and written as
\begin{equation}
  g^I_{ijkl}(z,\bar z)\pm \sum_{J} M^I_J g^J_{kjil}(1-z,1-\bar z)=0
\end{equation}
for all $I$, where $M_J^I$ is a matrix relating $T_{I,kjil}$ to $T_{J,ijkl}$. We can now apply functionals, specifically derivatives of the form $\partial^m \bar \partial^n$, on this equation in the usual manner.

The eventual result is an equation of the form
\begin{equation}
  \sum_\calO \vb \lambda^\intercal_\calO \cG_\calO^{\vb I} \vb \lambda_\calO = 0\,,
\end{equation}
where $\vb I$ is a new index combining both the four-point structure $I$ and the specific functional $(m,n)$ acting on the block, $\cG$ is a matrix constructed from the block, and $\vb
\lambda_\calO$ is a vector of nonvanishing OPE coefficients $\lambda^a_{ij\calO}$. We can rearrange this slightly by constructing the positive semidefinite matrices
\begin{equation}
  \mathcal P_\calO^{(a,ij);(b,kl)} = \lambda^a_{ij\calO} \lambda^b_{kl\calO} \succeq 0\,.
\end{equation}
After breaking the sum into internal, external, and identity parts, we obtain
\begin{equation}
  \label{eq:sdcrossing}
  \cG^{\vb I}_{\mathds{1}} + \mathrm{Tr}\,\left(\mathcal P_{ext}\cG^{\vb I}_{ext}\right)
  +\sum_{\rho}\sum_{\Delta}\mathrm{Tr}\,\left(\mathcal P_{\Delta,\rho}\cG^{\vb I}_{\Delta,\rho}\right) = 0
\end{equation}
for all $\vb I$. This is a semidefinite feasibility problem. We can impose spectrum assumptions by restricting the sum over $\Delta$ for various choices of $\rho$ in the third term, and can impose constraints on OPE coefficients by restricting $\mathcal P$.
Further details, including the specific choice of three- and four-point tensor structures, can be found in \cite{Erramilli:2022kgp}.

\subsection{Tiptop algorithm}

Placing an upper or lower bound on a scaling dimension is equivalent to extremizing the island in scaling dimension space.  We originally attempted this extremization via a navigator function \cite{Reehorst:2021ykw,Liu:2023elz}, but because the navigator turned out to be non-convex (and its Hessian ill-conditioned) in the external OPE coefficients, we elected to instead use a \texttt{tiptop} search \cite{Chester:2020iyt} combined with an OPE scan \cite{Kos:2016ysd}. The \texttt{tiptop} algorithm, which proceeds by iteratively refining an island for various values of the objective, is summarized below.

\SetArgSty{textup}
\begin{algorithm}[h] 
Let $\Dtx{feasible}$ be the largest value of  $\Dtx{obj}$ known to have a feasible point.

Let $\Dtx{ceiling}$ be the smallest value of  $\Dtx{obj}$ known not to have a feasible point.

Let $\Dtx{previous} = 0$.

\While{ $\Dtx{feasible} - \Dtx{previous} < \text{tolerance}$ }{
  Compute several allowed points at $\Dtx{obj} = \Dtx{feasible}$.

  Rescale the local coordinates via SVD of the island points.

  Continue computing points using adaptive mesh refinement to obtain the island boundary. This proceeds until the minimum feature size of the mesh is less than $f_\text{cutoff}$ times the smallest bounding box dimension of the allowed points.

  Let $\Dtx{previous} = \Dtx{feasible}$.

  Perform a binary search to find the largest $\Dtx{obj}$ between $\Dtx{previous}$ and
  $\Dtx{ceiling}$ such that the center of the island is still allowed. Let that be the new value of $\Delta_{\text{feasible}}$.}

Return $\Dtx{feasible}$.

\ 

  \caption{\texttt{tiptop} algorithm for maximizing a given $\Delta_\text{obj}$. For minimization, $\Delta_\text{obj}$ should be replaced with $-\Delta_\text{obj}$.}
\label{alg:tiptop}
\end{algorithm}

This algorithm is relatively simple, is guaranteed to converge for convex islands, and most importantly for our purposes, allows us to perform an OPE scan as part of the feasibility search. Incorporating the OPE information allows us to place much tighter bounds on $\Dsp$ and $\Dep$ and establish the irrelevance of both. In this work, we run \texttt{tiptop} with an $f_\text{cutoff}$ of $2$ and a stopping tolerance of $10^{-3}$.

This work uses the following additional software packages for computing bootstrap bounds:
\begin{itemize}
    \item \texttt{SDPB} (\url{https://github.com/davidsd/sdpb}), a semidefinite program solver designed for bootstrap problems \cite{Simmons-Duffin:2015qma,Landry:2019qug}.
    \item \texttt{blocks\_3d} (\url{https://gitlab.com/bootstrapcollaboration/blocks\_3d}), a C++ program for computing 3d conformal blocks \cite{Erramilli:2019njx,Erramilli:2020rlr}.
    \item \texttt{hyperion} (\url{https://github.com/davidsd/hyperion}), a Haskell library for managing high performance computing jobs.
    \item  \texttt{hyperion-bootstrap} (\url{https://gitlab.com/davidsd/hyperion-bootstrap}), a Haskell library for setting up bootstrap problems, along with several associated libraries (see appendix A of \cite{Erramilli:2022kgp}).
    \item \texttt{quadratic-net} (\url{https://gitlab.com/davidsd/quadratic-net}), a library for solving low-dimensional quadratically constrained quadratic programs via semidefinite relaxation, which is used in the OPE search~\cite{Chester:2019ifh}.
\end{itemize}

\section{Results and discussion} \label{sec:results}

\begin{table}[t!] 
\begin{center}
\renewcommand{\arraystretch}{1.1}
\resizebox{\columnwidth}{!}{
\begin{tabular}{ l|l l|l l }
  \hline\hline $N$~& $\Dsp$ lower bound & $\Dsp$ upper bound
  & $\Dep$ lower bound & $\Dep$ upper bound\\
  \hline
  2 & 3.071 & 3.328 & 3.169 & 3.835\\
  4 & 3.241 & 3.506 & 3.137 & 3.783\\
  8 & 3.189 & 3.650 & 3.167 & 3.842\\
\hline\hline 
\end{tabular}}
\end{center}
\caption{Computed $\Lambda=19$ bootstrap results. All bounds assume $\Delta_{\sigma_T} > 2$, $\Delta_{\psi'} > 2$, and $\Delta_{\chi} > 3.5$. Additionally, bounds on $\Dsp$ assume that $\Delta_{\sigma''}>4.5$ and $\Dep>3$, while bounds on $\Dep$ assume that $\Delta_{\epsilon''}>4$ and $\Dsp>2.5$.}
\label{tab:bounds}
\end{table}

\begin{figure}[th]
  \centering
    \includegraphics[width=.85\textwidth]{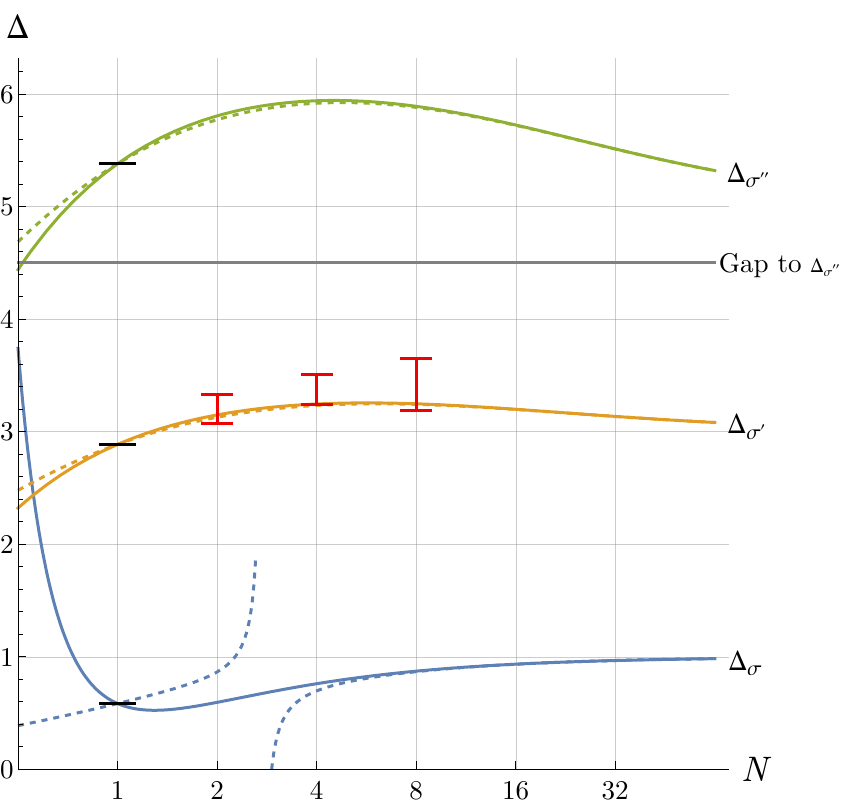}
  \caption{Bounds on $\Dsp$ (red) at $\Lambda=19$, plotted against Pad\'e approximants for the parity-odd operators. Solid lines indicate (2, 1) Padé approximants, while dashed lines indicate (1, 2) approximants.  The boundary conditions at $N=1$ corresponds to the $\mathcal N = 1$ super-Ising bootstrap data.}
  \label{fig:boundssig}
\end{figure}

\begin{figure}[th]
    \centering
    \includegraphics[width=.85\textwidth]{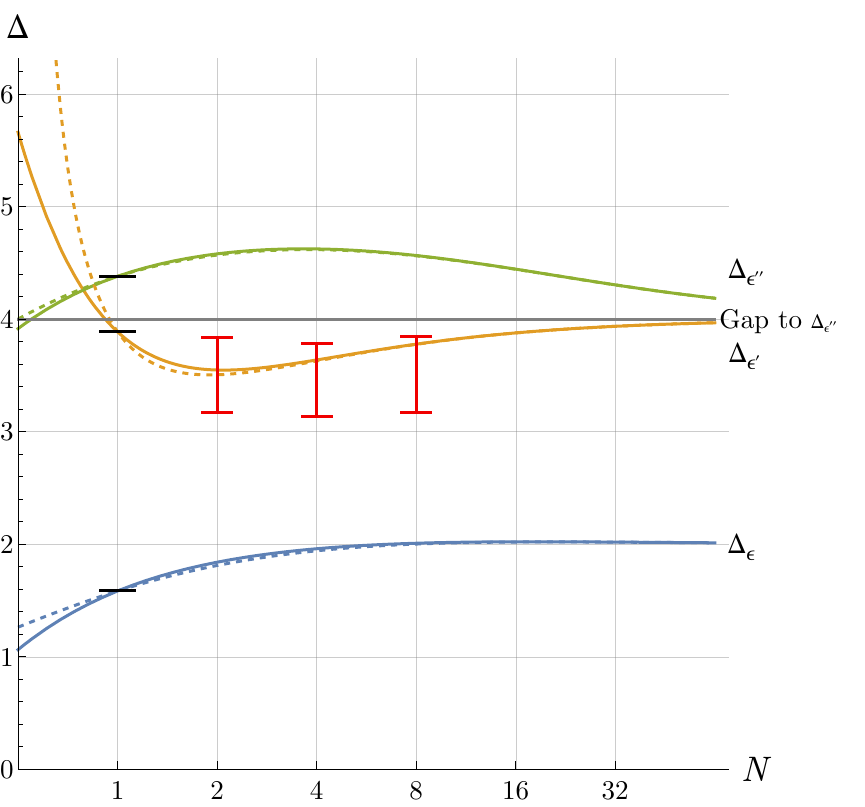}
  \caption{Bounds on $\Dep$ (red) at $\Lambda=19$, plotted against Pad\'e approximants for the parity-even operators. Solid lines indicate (2,1) Padé approximants, while dashed lines indicate (1,2) approximants.  The boundary conditions at $N=1$ corresponds to the $\mathcal N = 1$ super-Ising bootstrap data.}
  \label{fig:boundseps}
\end{figure}

\begin{figure}[th] 
  \centering
  \includegraphics[width=.74\textwidth]{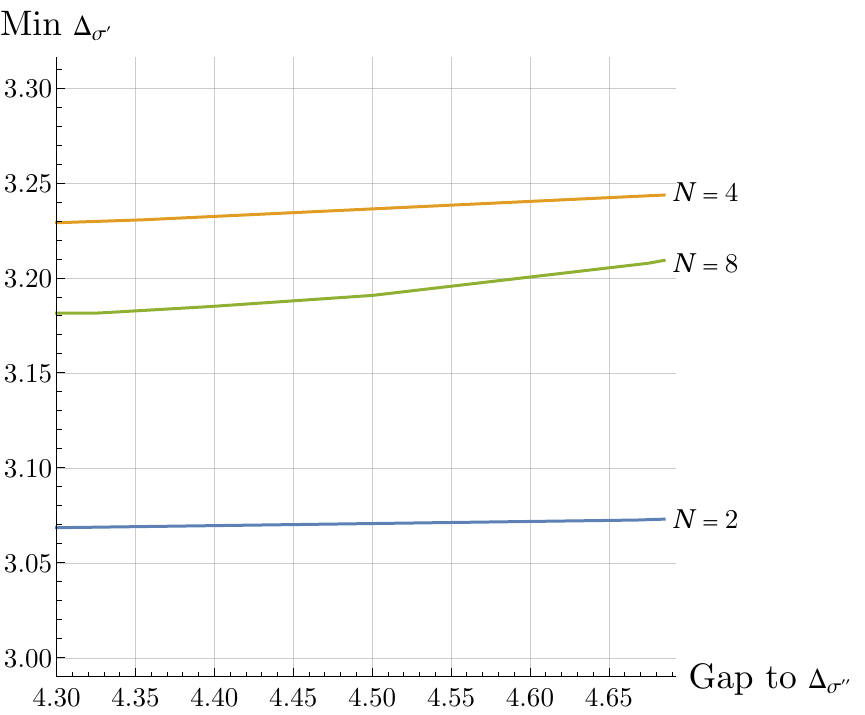}
  \caption{Sensitivity of the $\Delta_{\sigma'}$ lower bound to different $\Delta_{\sigma''}$ gaps at $\Lambda=19$, for fixed $(\Delta_\psi,\Delta_\sigma,\Delta_\eps)$.}
  \label{fig:sensSig}
\end{figure}

\begin{figure}[th] 
  \centering
  \includegraphics[width=.74\textwidth]{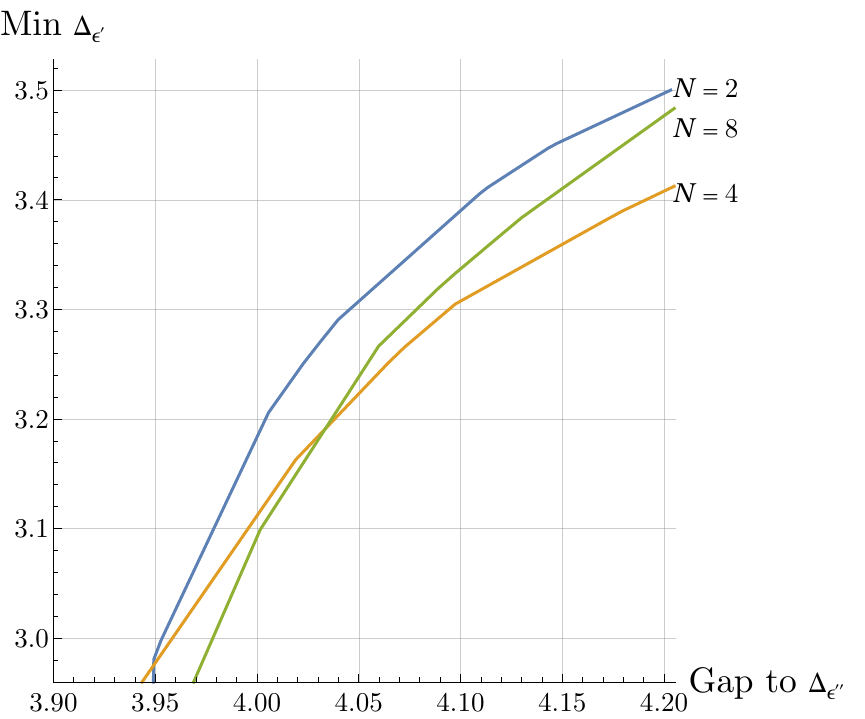}
  \caption{Sensitivity of the $\Delta_{\eps'}$ lower bound to different $\Delta_{\eps''}$ gaps at $\Lambda=19$, for fixed $(\Delta_\psi,\Delta_\sigma,\Delta_\eps)$.}
  \label{fig:sensEps}
\end{figure}

Bootstrap results for $\Dsp$ and $\Dep$, computed at derivative order $\Lambda=19$, are shown in table \ref{tab:bounds} and plotted against Padé approximants in figures~\ref{fig:boundssig} and~\ref{fig:boundseps}. Our results establish that $\Dsp$ and $\Dep$ are irrelevant at $N=2$, $4$, and $8$, given very conservative gap assumptions on $\Delta_{\sigma''}$ and $\Delta_{\epsilon''}$. In addition to constraining the possible RG flows that can be triggered by singlet scalar deformations, this places the bootstrap bounds computed in \cite{Erramilli:2022kgp} on a much stronger footing.

We have also checked the sensitivity of our bounds to the gap assumptions. We have found that our $\Delta_{\sigma'}$ bounds are very insensitive to our assumption on $\Delta_{\sigma''}$. Figure \ref{fig:sensSig} illustrates the sensitivity of the $\Dsp$ lower bound with varying gap assumptions on $\Delta_{\sigma''}$, for the fixed values of $(\Delta_\psi,\Delta_\sigma,\Delta_\eps)$ that saturate the lower bound on $\Dsp$ in table \ref{tab:bounds}.  For all $N$, lowering the gap to $\sim 4.3$ was insufficient to take $\Dsp$ below marginality.\footnote{We should note that this is not perfectly rigorous, as the point in $(\Delta_\psi,\Delta_\sigma,\Delta_\eps)$-space that minimizes $\Dsp$ may change slightly depending on the gap assumptions, but it serves as a very easy sanity check.} This is less true for $\Dep$ (see figure \ref{fig:sensEps}), where if $\Delta_{\eps''}$ becomes sufficiently low the $\eps''$ operator can effectively play the role of $\eps'$ and allow $\eps'$ to decouple. However, given how conservative our initial gap of $\Delta_{\eps''} \ge 4$ is compared with the Pad\'e estimates, we are confident that the physical $\eps'$ operator is irrelevant.

In the future, it would be interesting to examine the full RG flow structure of the $O(N)$ GNY CFTs and related fixed points for various $N$. It will also be important to understand how to successfully implement the navigator~\cite{Reehorst:2021ykw} and skydiving~\cite{Liu:2023elz} approaches to the rigorous study of the subleading spectra of these theories. There are additionally many other interesting fermionic theories, such as the chiral Ising, chiral XY, and chiral Heisenberg models, whose interactions preserve a subgroup of the $O(N)$ flavor symmetry and are relevant to various condensed matter systems~\cite{Zerf:2017zqi, Erramilli:2022kgp}. Bootstrap studies of these theories are a logical next step in understanding the space of fermionic 3d CFTs. Ultimately, these models serve as elegant testing grounds for the development of general bootstrap methods which can eventually be applied to map out the full space of CFTs and their physical deformations.

\newpage
\section*{Acknowledgements}

We thank Vasiliy Dommes, Rajeev Erramilli, Mark Gonzalez, Luca Iliesiu, Petr Kravchuk, Aike Liu, Marten Reehorst, Slava Rychkov, and David Simmons-Duffin for discussions. The authors were supported by Simons Foundation grant 488651 (Simons Collaboration on the Nonperturbative Bootstrap) and DOE grant DE-SC0017660. This work used the Expanse cluster at the San Diego Supercomputing Center (SDSC) through allocation PHY190023 from the Advanced Cyberinfrastructure Coordination Ecosystem: Services \& Support (ACCESS) program, which is supported by National Science Foundation grants \#2138259, \#2138286, \#2138307, \#2137603, and \#2138296. Computations were also performed on the Yale Grace computing cluster, supported by the facilities and staff of the Yale University Faculty of Sciences High Performance Computing Center.

\bibliographystyle{JHEP}
\bibliography{refs}

\end{document}